\documentclass[twocolumn,showpacs,preprintnumbers,amsmath,amssymb]{revtex4}

\usepackage{graphicx}
\usepackage{dcolumn}
\usepackage{bm}
\usepackage{epsfig}
\usepackage{epsfig}
\usepackage{amsmath}
\usepackage{amsfonts}

\newcommand{\ba}{\begin{eqnarray}}
\newcommand{\ea}{\end{eqnarray}}
\newcommand{\be}{\begin{equation}}
\newcommand{\ee}{\end{equation}}
\newcommand{\bea}{\begin{eqnarray}}
\newcommand{\eea}{\end{eqnarray}}

\newcommand{\unit}{\mathbb{I}}

\newcommand{\ten}{\otimes}

\usepackage{theorem}

\theorembodyfont{\rmfamily}
\newtheorem{remark}{Remark}
\theoremstyle{break}

\def\QED{~\rule[-1pt]{5pt}{5pt}\par\medskip}

\DeclareMathOperator{\tr}{tr}

\def\c {\mathfrak{c}}

\def\E{\mathcal{E}}
\def\H{\mathcal{H}}
\def\Ai{\A_{\rm in}}
\def\Ao{\A_{\rm out}}
\def\B{\mathcal{B}}
\def\Bi{\B_{\rm in}}
\def\Bo{\B_{\rm out}}
\def\A{\mathcal{A}}

\def\ha{\mathcal{H}_A}
\def\hb{\mathcal{H}_B}
\def\s{\hat{S}}
\def\a{\hat{A}}
\def\S{\mathcal{P}}
\def\c{C^d}
\def\t{\mbox{tr}}
\def\i{I}

\newcommand{\ket}[1]{\mbox{$| #1 \rangle$}}
\newcommand{\bra}[1]{\mbox{$\langle #1 |$}}

\begin{document}


\title{Classification of the separable maps which preserve Werner states }

\author{$^1$Haidong Yuan\footnotemark[1],$^2$Lluis Masanes}

\affiliation{ $^1$Massachusetts Institute of Technology, Cambridge, MA 02139\\
$^2$ICFO-Institut de Ciencies Fotoniques, 08860 Castelldefels (Barcelona), Spain}

\date{\today}

\begin{abstract}
We classify the completely-positive maps acting on two $d$-dimensional systems which commute with all $U\otimes U$ unitaries, where $U\in SU(d)$. This set of operations map Werner states to Werner states. We find a simple condition for a map being implementable by stochastic local operations and classical communication (SLOCC). We show that all PPT-preserving maps can be implemented by SLOCC. This can be used to prove that the entanglement of Werner states cannot be stochastically increased, even if we allow PPT entanglement for free.
\end{abstract}


\maketitle
\footnotetext[1]{$^1$haidong@mit.edu,$^2$masanes@damtp.cam.ac.uk}
\section{Introduction}
Entanglement play a crucial role in the application of quantum
information science, such as quantum key distribution
\cite{Ekt91}, superdense coding \cite{Ben922}, quantum
teleportation \cite{Ben93} and quantum error correction
\cite{QECC} etc. Among several bipartite entangled states, Werner
states~\cite{Werner:1989} provide the simplest example of  mixed
states possessing entanglement and they play an important role in
entanglement purification \cite {Ben96}, nonlocality
\cite{Werner:1989}, entanglement measures \cite{Shor01}, etc.
Werner states are a useful family of states depending on a single
parameter, more theoretical investigations on Werner states can be
seen in \cite{Wer}.

On the other hand, it is often of interest in quantum information
processing to determine if a given state can be transformed to
some other desired state by local operations. Indeed,
convertibility between two (entangled) states using local quantum
operations assisted by classical communication (LOCC) is closely
related to the problem of quantifying the entanglement associated
to each quantum system. Intuitively, one expects that a (single
copy) entangled state can be locally and deterministically
transformed to a less entangled one but not the other way round.
In this paper we consider transformations whose success
probability can be smaller than one, but of course, it has to be
strictly larger than zero. This set of transformations is called
{\em stochastic local operations and classical communication}
(SLOCC), then in general it is possible to transform the state the
other way around, i.e., transform a less entangled state to a more
entangle state with some success probability. However in this
paper, by studying the separable maps that transfer from Werner
states into Werner states, we show that the entanglement of Werner
states can not increase under these maps even they can be
stochastic local operations and classical communication(SLOCC).



The remainder of the paper is organized as follows: In following
two sections, we give a brief introduction of the Werner states
and separable maps. We then give our classification of the
separable maps that preserve the set of Werner states, which is,
by Jamiolkowsky theorem, equivalent to the classification of a
class of symmetric states. in section \ref{sect:application} we
present an application of this classification.

\section{Werner States}

Consider a tensor-product Hilbert space $\ha\otimes\hb$, where
$\ha=\hb=\c$.
The swap unitary matrix $F$ is defined as
$F\ket{\psi}\otimes\ket{\phi} =\ket{\phi}\otimes\ket{\psi}$ for
all $\ket{\psi},\ket{\phi}\in\ha$. The symmetric and antisymmetric
projectors are respectively
\begin{eqnarray}
  S &=& \frac{\i+F}{2}, \\
  A &=& \frac{\i-F}{2},
\end{eqnarray}
where $I$ is the identity matrix acting on $\ha$. Analogously the
symmetric and antisymmetric normalized states are $\s=S/\t S$ and
$\a=A/\t A$.

Werner states, denoted $\omega_\nu$, are the ones that commute
with all unitaries of the form $U\otimes
U$\cite{Werner:1989,Werner:2001}, where $U$ acts on $\c$, and can
be written as
\begin{equation}\label{}
    \omega_{\nu} = \nu \a + (1-\nu) \s,
\end{equation}
where $\nu\in[0,1]$. We define the depolarization map as
\begin{equation}\label{depolarization}
    \Lambda[\rho]= \a\,\t\rho A + \s\,\t\rho S.
\end{equation}
which is just projection to the Werner states. This map can be
physically implemented by LOCC in the following way
\begin{equation}\label{dep}
    \Lambda[\rho] = \int dU\
    U\otimes U\, \rho\ (U\otimes U)^\dag ,
\end{equation}
where $dU$ is the Haar measure over $SU(d)$. It is known that
$\omega_\nu$ is separable for $\nu\in[0,1/2]$ and entangled for
$\nu\in(1/2,1]$~\cite{Werner:1989,Werner:2001}.

\section{Separable maps and SLOCC}
In this section, we give an brief introduction of the separable
maps and SLOCC, we follow the treatment in ~\cite{Liang}. To begin
with, a separable complete positive map(CPM), denoted by $\E_s$
takes the following
form~\cite{E.M.Rains:9707002,V.Vedral:PRA:1998}
\begin{equation}\label{Eq:SeparableMap}
    \E_s:\rho\to\sum_{i=1}^n(A_i\ten B_i)~\rho~(A_i^\dag\ten B_i^\dag),
\end{equation}

where $\rho$ acts on $\H_{\Ai}\ten\H_{\Bi}$, $A_i$ acts on
$\H_{\Ai}$, $B_i$ acts on $\H_{\Bi}$. If, moreover,
\begin{equation}\label{Eq:TracePreserving}
    \sum_i \left(A_i\ten{B_i}\right)^\dag
    \left(A_i\ten{B_i}\right)=\unit,
\end{equation}
the map is trace-preserving, i.e., if $\rho$ is normalized, so is
the output of the map $\E_s(\rho)$. Equivalently, the
trace-preserving condition demands that the transformation from
$\rho$ to $\E_s(\rho)$ can always be achieved with certainty. It
is well-known that all LOCC transformations are of the form
Eq.~\eqref{Eq:SeparableMap} but the converse is not
true~\cite{C.H.Bennett:PRA:1999}.

However, if we allow the map $\rho\to\E_s(\rho)$ to fail with some
probability $p<1$, the transformation from $\rho$ to $\E_s(\rho)$
can always be implemented probabilistically via LOCC. In other
words, if we do not impose Eq.~\eqref{Eq:TracePreserving}, then
Eq.~\eqref{Eq:SeparableMap} represents, up to some normalization
constant, the most general LOCC possible on a bipartite quantum
system. These are the SLOCC transformations~\cite{W.Dur:PRA:2000}.

Let us also recall the Choi-Jamio{\l}kowski isomorphism~\cite{J}
between CPM and quantum states: for every (not necessarily
separable) CPM $\E:\H_{\Ai}\ten\H_{\Bi}\to\H_{\Ao}\ten\H_{\Bo}$
there is a unique -- again, up to some positive constant $\alpha$
-- quantum state $\rho_\E$ corresponding to $\E$:
\begin{equation}\label{Eq:JamiolkowskiState}
    \rho_\E=\alpha~\E\otimes \mathcal{I} \left(\ket{\Phi^+}_{\Ai}\bra{\Phi^+}\ten
    \ket{\Phi^+}_{\Bi}\bra{\Phi^+}\right),
\end{equation}
where
$\ket{\Phi^+}_{\Ai}\equiv\sum_{i=1}^{d_{\Ai}}\ket{i}\ten\ket{i}$
is the unnormalized maximally entangled state of dimension
$d_{\Ai}$ (likewise for $\ket{\Phi^+}_{\Bi}$). In
Eq.~\eqref{Eq:JamiolkowskiState}, it is understood that $\E$ only
acts on half of $\ket{\Phi^+}_{\Ai}$ and half of
$\ket{\Phi^+}_{\Bi}$. Clearly, the state $\rho_\E$ acts on a
Hilbert space of dimension $d_{\Ai}\times d_{\Ao}\times
d_{\Bi}\times d_{\Bo}$, where $d_{\Ao}\times d_{\Bo}$ is the
dimension of $\H_{\Ao}\ten\H_{\Bo}$.

Conversely, given a state $\rho_\E$ acting on
$\H_{\Ao}\ten\H_{\Bo}\ten\H_{\Ai}\ten\H_{\Bi}$, the corresponding
action of the CPM $\E$ on some $\rho$ acting on
$\H_{\Ai}\ten\H_{\Bi}$ reads:
\begin{equation}\label{Eq:State->CPM}
    \E(\rho)=\frac{1}{\alpha}\tr_{\Ai\Bi}\left[\rho_\E\left(\unit_{\Ao\Bo}\ten\rho^{\mbox{\tiny T}}\right)\right],
\end{equation}
where $\rho^{\mbox{\tiny T}}$ denote transposition of $\rho$ in
some local bases of $\H_{\Ai}\ten\H_{\Bi}$. For a trace-preserving
CPM, it then follows that we must have
$\tr_{\Ao\Bo}(\rho_\E)=\alpha\unit_{\Ai\Bi}$. A point that should
be emphasized now is that $\E$ is a separable map [c.f.
Eq.~\eqref{Eq:SeparableMap}] if and only if the corresponding
$\rho_\E$ given by Eq.~\eqref{Eq:JamiolkowskiState} is separable
across $\H_{\Ai}\ten\H_{\Ao}$ and
$\H_{\Bi}\ten\H_{\Bo}$~\cite{J.I.Cirac:PRL:2001}. Moreover, at the
risk of repeating ourselves, the map $\rho\to\E(\rho)$ derived
from a separable $\rho_\E$ can always be implemented locally,
although it may only succeed with some (nonzero) probability.
Hence, if we are only interested in transformations that can be
performed locally, and not the probability of success in mapping
$\rho\to\E(\rho)$, the normalization constant $\alpha$ as well as
the normalization of $\rho_\E$ becomes irrelevant. This is the
convention that we will adopt for the rest of this section.

\section{Symmetric maps}

Let us consider completely positive maps that transform Werner
states into Werner states in $\ha\otimes\hb$. By using the
Jamiolkowsky theorem~\cite{J}, each of this maps $\E$ has
associated a state $\rho_{\E}$ acting on $\ha\otimes\hb \otimes
\mathcal{H}_{A'} \otimes \mathcal{H}_{B'}$, where also
$\mathcal{H}_{A'} = \mathcal{H}_{B'} = \c$. The action of the map
$\E$ associated to $\rho_{\E}$ is the following
\begin{equation}\label{}
    \E[\rho]=\t_{AB}
    \left( \rho_{\E}(I_{A'B'}\ten\rho^T ) \right) .
\end{equation}
Because of this rule, we refer to $\ha\otimes\hb$ as the input
space, and to $\mathcal{H}_{A'} \otimes \mathcal{H}_{B'}$ as the
output space. One can see that the states $\rho_{\E}$ associated
to these maps commute with all unitaries of the
form~\cite{Werner:1989,Werner:2001}
\begin{equation}\label{group}
    U\otimes U \otimes V\otimes V,
\end{equation}
where $U$ and $V$ act on $\c$. From the Jamiolkowsky theorem we
know that a map $\E$ is separable if, and only if, its associated
state $\rho_{\E}$ is separable across $\ha\ten\mathcal{H}_{A'}$
and $\hb\ten\mathcal{H}_{B'}$~\cite{J.I.Cirac:PRL:2001}. In the
following we classify all separable states of this kind.

The states that commute with the group (\ref{group}) are of the
form~\cite{Werner:1989,Werner:2001}
\begin{equation}\label{state}
    \xi= \lambda_1\a\otimes\a + \lambda_2\a\otimes\s +
    \lambda_3\s\otimes\a + \lambda_4\s\otimes\s ,
\end{equation}
with $\lambda_i\geq 0$ and $\sum_i \lambda_i =1$, where the tensor
is across subspace $AB$ and $A'B'$. In the following we specify
states with four-dimensional vectors $\vec{\lambda}$. Let us see
that the following five states are separable.
\begin{eqnarray}
  \vec{\lambda}^{(1)} &=& \left(0,0,0,1\right) \\
  \vec{\lambda}^{(2)} &=& \left(0,\frac{1}{2},0,\frac{1}{2}\right) \\
  \vec{\lambda}^{(3)} &=& \left(0,0,\frac{1}{2},\frac{1}{2}\right) \\
  \vec{\lambda}^{(4)} &=& \left(\frac{1}{4},\frac{1}{4},
  \frac{1}{4},\frac{1}{4}\right) \\
  \vec{\lambda}^{(5)} &=& \left(\frac{1}{2}-\frac{1}{2d},0,
  0,\frac{1}{2}+\frac{1}{2d}\right)
\end{eqnarray}
The states $\vec{\lambda}^{(1)}, \ldots \vec{\lambda}^{(4)}$ are
separable because they are products of the two separable Werner
states $\omega_0$ and $\omega_{1/2}$. The state
$\vec{\lambda}^{(5)}$ is separable because it is the output of the
LOCC map
\begin{equation}\label{}
    \Delta[\rho]= \int\! dU\! \int\! dV\,
    U\otimes U\otimes V\otimes V \rho\
    (U\otimes U\otimes V\otimes V)^\dag
\end{equation}
when the input is the product state $\sum_{ks=1}^d
\ket{ks}_{AB}\otimes\ket{ks}_{A'B'}/d$. The map $\Xi$ associated
to the state $\vec{\lambda}^{(5)}$ is precisely the depolarization
map (\ref{depolarization},\ref{dep}).

Let us denote by $\S$ the convex polytope generated by
$\vec{\lambda}_{1}, \ldots \vec{\lambda}_5$. Clearly, all points
in $\S$ correspond to separable states. Let us see that only these
states are the separable ones. The set $\S$ can be characterized
by a finite number of linear inequalities of the form
\begin{equation}\label{ineq}
    \vec{\lambda} \cdot \vec{\mu} \geq 0.
\end{equation}
We can chose the constant in the right-hand side of the
inequalities to be zero because all the points in $\S$ are inside
the normalization hyperplane. Some of the inequalities $\vec{\mu}$
correspond to the positivity conditions of $\vec{\lambda}$, the
rest (the relevant ones) are
\begin{eqnarray}\label{mus}
  \vec{\mu}^{(1)} &=& \left( 1,-1,-1,1 \right), \\
  \vec{\mu}^{(2)} &=& \left( -d-1,d+1,-d+1,d-1 \right), \\
  \vec{\mu}^{(3)} &=& \left( -d-1,-d+1,d+1,d-1 \right).
\end{eqnarray}
The vectors $\vec{\mu}^{(i)}$ are the extreme points of the dual
polytope of $\S$. In general one can find them with standard
software, but this case is simple enough for doing it by hand.

The hermitian matrices corresponding to the vectors $\vec{\mu}$
are
\begin{equation}\label{w-m}
    W_{\vec{\mu}} = \mu_1 A\otimes A +\mu_2 A\otimes S
    +\mu_3 S\otimes A +\mu_4 S\otimes S.
\end{equation}
Notice that in order to make
\begin{equation}\label{}
    \t\, \xi W = \vec{\lambda}\cdot\vec{\mu}
\end{equation}
we must write projectors $A$ and $S$ instead of normalized states
$\a$ and $\s$. Up to an unimportant proportionality factor we can
be write the hermitian matrices associated to (\ref{w-m}) as
\begin{eqnarray}
  W^{(1)} &=& F\otimes F, \\
  W^{(2)} &=& d I\otimes F - F\otimes F, \\
  W^{(3)} &=& d F\otimes I - F\otimes F.
\end{eqnarray}
In what follows, we prove that these three operators are
entanglement witnesses, that is, their expected values with
product states are non-negative:
\begin{equation}\label{w}
    \bra{\alpha}\otimes\bra{\beta}\ W^{(i)}\ \ket{\alpha}\otimes\ket{\beta}
    \geq 0
\end{equation}
for all $\ket{\alpha}\in \ha\ten\mathcal{H}_{A'}, \ket{\beta} \in
 \hb\otimes \mathcal{H}_{B'}$. For doing this the
following identities are useful,
\begin{eqnarray}
  \nonumber
  \bra{\alpha_{AA'} \beta_{BB'}} I\otimes F
  \ket{\alpha_{AA'} \beta_{BB'}} &=&
  \t \left( \alpha^\dag \alpha \beta^\dag \beta \right), \\
  \nonumber
  \bra{\alpha_{AA'} \beta_{BB'}} F\otimes I
  \ket{\alpha_{AA'} \beta_{BB'}} &=&
  \t \left( \alpha \alpha^\dag \beta \beta^\dag \right), \\
  \nonumber
  \bra{\alpha_{AA'} \beta_{BB'}} F\otimes F
  \ket{\alpha_{AA'} \beta_{BB'}} &=&
  \left|\, \t\, \alpha \beta^\dag  \right|^2,
\end{eqnarray}
where the matrix $\alpha_{ij}$ is defined as $\ket{\alpha_{AA'}}=
\sum_{ij} \alpha_{ij} \ket{i}_A\otimes\ket{j}_{A'}$, and
analogously for $\beta$. From the last equality, one can
straightforwardly see that $W^{(1)}$ is a witness. To prove that
$W^{(2)}$ is also a witness we define the matrix $\gamma=\alpha
\beta^\dag$ and write the expected with an arbitrary product state
as
\begin{equation}\label{min}
    \bra{\alpha \beta} W^{(2)} \ket{\alpha \beta}=
    d\, \t \gamma^\dag \gamma - \left|\t\gamma\right|^2\geq 0.
\end{equation}
The inequality comes from Cauchy-Schwarz inequality as we can
write $d$ as $\t\i^\dag\i$ and $\t\gamma$ as $\t\i\gamma$.
Similarly, one can prove that $W^{(3)}$ is an entanglement
witness.

The fact that $W^{(1)}$, $W^{(2)}$ and $W^{(3)}$ are witnesses
implies that all points outside the polytope $\S$ are
nonseparable. Concluding, $\S$ is the set of separable symmetric
states.

\remark{ One can see that the partial transpositions of $S$ and
$A$ are
\begin{eqnarray}
  S^\Gamma &=& \frac{\i + d \Phi}{2}, \\
  A^\Gamma &=& \frac{\i - d \Phi}{2},
\end{eqnarray}
where $\Phi$ stands for the projector onto the maximally entangled
state $\ket{\Phi}=\sum_{d=1}^d \ket{kk}$. If one performs the
partial transposition on the generic state (\ref{state}), and
imposes that the result is positive semi-definite, one obtains the
three inequalities associated to $\vec{\mu}^{(1)},
\vec{\mu}^{(2)}, \vec{\mu}^{(3)}$, in (\ref{mus}). This means that
the set of symmetric PPT(positive partial transpose) states is
also $\S$, which means the set of PPT preserve maps which preserve
Werner states coincide with the set of separable maps which
preserve the Werner states. So we actually also classified the PPT
preserve maps which preserve Werner states.
%
 }

\section{Applications}
\label{sect:application}

We can use our classification of separable symmetric maps to prove
that the entanglement of Werner states cannot increase under
stochastic local operations and classical communication (SLOCC).
It is known that all separable maps can be physically implemented
with SLOCC, and obviously, all SLOCC are separable maps. Then,
what we want to prove is that, if
\begin{equation}
    \Xi[\nu\a + (1-\nu)\s] = \nu'\a + (1-\nu')\s,
\end{equation}
where $\nu\geq 1/2$ and $\Xi$ is a separable map, then
$\nu'\leq\nu$.

By using the Jamiolkowsky isomorphism we have that
\begin{eqnarray}\label{}
\aligned
    \Xi[\nu\a + (1-\nu)\s] &=\\
     2 \frac{\nu}{d(d-1)}\left[\lambda_1\a+\lambda_2\s\right]
    &+2 \frac{1-\nu}{d(d+1)}\left[\lambda_3\a+\lambda_4\s\right]
    \nonumber.
    \endaligned
\end{eqnarray}
The new value of the parameter is
\begin{equation}\label{}
    \nu'=\frac
    {2\frac{\nu}{d(d-1)}\lambda_1 + 2\frac{1-\nu}{d(d+1)}\lambda_3}
    {2 \frac{\nu}{d(d-1)}\left[\lambda_1+\lambda_2\right]
    +2 \frac{1-\nu}{d(d+1)}\left[\lambda_3+\lambda_4\right]}
\end{equation}
Now, let us assume $\nu'>\nu$. After some algebra we transform
$\nu'>\nu$ into
\begin{equation}\label{}
    (d-1)\left[ \eta\lambda_4 - \eta^2\lambda_3 \right] +
    (d+1)\left[ \lambda_2 - \eta\lambda_1 \right] < 0,
\end{equation}
where $\eta=(1-\nu)/\nu$. Because $0<\eta<1$ then
$-\eta\lambda_3\leq-\eta^2\lambda_3$ and
$\eta\lambda_2\leq\lambda_3$. Therefore, if the above inequality
is true we have
\begin{equation}\label{}
    (d-1)\left[ \lambda_4 - \lambda_3 \right] +
    (d+1)\left[ \lambda_2 - \lambda_1 \right] < 0,
\end{equation}
which is precisely $\vec{\lambda}\cdot \vec{\mu}^{(2)} < 0$. This
is in contradiction with the fact that $\Xi$ is separable,
therefore $\nu'\leq \nu$ must hold.
%
%
%
%
%
%
%
%
%
%

\section{Conclusions}
In summary, we have explicitly characterized the complete positive
maps acting on two $d$-dimensional systems which commute with all
$U\otimes U$ unitaries. A simple condition is also given on those
maps which can be implemented by SLOCC. We achieved this by using
the Jamiolkowsky theorem: instead of characterizing the complete
positive maps directly, we equivalently characterized the states
of four $d$-dimensional systems which commute with all $U\otimes
U\otimes V\otimes V$ unitaries, where $U,V\in SU(d)$. This enabled
us giving conditions on the separable and PPT preserve maps which
preserve the Werner states. With these conditions, we showed that
the entanglement of Werner states cannot be increased under SLOCC.

The fact that the entanglement of Werner states cannot be
increased under SLOCC has been the key to prove that each
bipartite entangled state can increase the teleportation power of
another state \cite{m}. Establishing that all entangled states are
useful for quantum information processing.

\medskip {\bf Acknowledgements.} The authors are thankful to Andrew Doherty for valuable comments. This work has been financially supported by the EU project QAP (IST-3-015848), the spanish MEC (FIS2005-04627, FIS2007-60182, Consolider QOIT), and Caixa Manresa.

\end{document}